\documentclass[11pt]{article}
\usepackage{amsmath,amssymb,amsfonts,epsf,graphicx,color}
\usepackage[nosort]{cite}
\usepackage[margin=1in]{geometry}

\makeatletter \@addtoreset{equation}{section} \makeatother

\newcommand{\fft}[2]{{\frac{#1}{#2}}}
\newcommand{\ft}[2]{{\textstyle\frac{#1}{#2}}}

\let\a=\alpha

\def\nn{\nonumber}

\newcommand{\be}{\begin{equation}}
\newcommand{\ee}{\end{equation}}
\def\ba{\begin{array}}
\def\ea{\end{array}}
\def\ft#1#2{{\textstyle{\frac{\scriptstyle #1}{\scriptstyle #2}}}}
\def\fft#1#2{\frac{#1}{#2}}

\def\sst#1{{\scriptscriptstyle #1}}

\def\td{\tilde}

\def\dalemb#1#2{{\vbox{\hrule height .#2pt
        \hbox{\vrule width.#2pt height#1pt \kern#1pt
                \vrule width.#2pt}
        \hrule height.#2pt}}}

\newcommand{\bea}{\begin{eqnarray}}
\newcommand{\eea}{\end{eqnarray}}

\newcommand{\Tr}{\mbox{Tr}}
\def\0{{\sst{(0)}}}
\def\1{{\sst{(1)}}}
\def\2{{\sst{(2)}}}
\def\3{{\sst{(3)}}}
\def\4{{\sst{(4)}}}
\def\5{{\sst{(5)}}}
\def\6{{\sst{(6)}}}
\def\7{{\sst{(7)}}}
\def\8{{\sst{(8)}}}

\def\ep{{\epsilon}}

\def\R{\rlap{\rm I}\mkern3mu{\rm R}}

\def\R{\rlap{\rm I}\mkern3mu{\rm R}}

\def\R{{{\mathbb R}}}

\def\CP{{{\mathbb C}{\mathbb P}}}

\begin{document}

\begin{center}\ \\ \vspace{50pt}
{\Large {\bf Marginal Deformations of Non-Relativistic Field Theories}}\\ 
\vspace{30pt}

Davron Mallayev$^{1}$, Justin F. V\'azquez-Poritz$^{1,2}$ and Zhibai Zhang$^{1,2}$
\vspace{20pt}

$^1${\it Physics Department\\ New York City College of Technology, The City University of New York\\ 300 Jay Street, Brooklyn NY 11201, USA}

\vspace{10pt}
$^2${\it The Graduate School and University Center, The City University of New York\\ 365 Fifth Avenue, New York NY 10016, USA}\\

\vspace{20pt}

{\tt davron.mallayev@mail.citytech.cuny.edu\\ jvazquez-poritz@citytech.cuny.edu\\ zzhang@citytech.cuny.edu}

\end{center}

\vspace{30pt}

\centerline{\bf Abstract}

We construct the supergravity duals of marginal deformations of a $(0,2)$ Landau-Ginsburg theory that describes the supersymmetric lowest Landau level. These deformations preserve supersymmetry and it is proposed that they are associated with the introduction of a phase in the $(0,2)$ superpotential. We also consider marginal deformations of various field theories that exhibit Schr\"odinger symmetry and Lifshitz scaling. This includes countably-infinite examples with dynamical exponent $z=2$ based on the Sasaki-Einstein spaces $Y^{p,q}$ and $L^{p,q,r}$, as well as an example with general dynamical exponent $z\ge 1$.

\thispagestyle{empty}

\pagebreak
%\voffset=0pt
%\setcounter{page}{1}

%\tableofcontents
%\addtocontents{toc}{\protect\setcounter{tocdepth}{3}}
%\newpage

%%%%%%%%%%%%%%%%%%%%%%%%%%%%%%%%%%%%%%%%

\section{Introduction}

The AdS/CFT correspondence provides a technique for studying certain strongly-coupled conformal field theories in terms of string theory on a weakly curved spacetime (see \cite{agmoo} for a review). In recent years, holographic techniques have been used to study strongly-coupled condensed matter systems, such as atomic gases at ultra-low temperature. Gravitational backgrounds that provide descriptions of these Lifshitz and Schr\"{o}dinger-like fixed points have been proposed in \cite{Kachru} and \cite{Son,Balasubramanian2}, respectively. In these systems, rather than obeying conformal scale invariance, the temporal and spatial coordinates scale anisotropically. Another example of a condensed matter system for which a dual gravitational description has been proposed is the effective field theory of the supersymmetric lowest Landau level, which is a $(0,2)$ Landau-Ginsburg model \cite{Almuhairi2}.

All of these systems can undergo marginal deformations, in the sense that the deformed theories preserve the non-relativistic symmetries of the undeformed theory. We will employ a solution-generating technique that is based on U-duality in order to find gravitational backgrounds in type IIB theory that correspond to marginal deformations of various non-relativistic theories. The procedure can be outlined as follows. Begin with a theory that has a global symmetry group that includes $U(1)\times U(1)$, which corresponds to a two-torus subspace of the type IIB background. T-dualize along one of the $U(1)$ directions to type IIA theory. Lifting the solution to eleven dimensions provides a third direction which is associated with a $U(1)$ symmetry. One can now apply an $SL(3,R)$ transformation along these $U(1)^3$ directions. Dimensionally reducing to type IIA theory and T-dualizing back to type IIB theory along the shifted directions yields a new type IIB solution.

This method for finding new solutions has been widely used, one instance of which was to generate the type IIB supergravity background that corresponds to marginal deformations of ${\cal N}=4$ super Yang-Mills theory \cite{Lunin}. In addition to the $U(1)\times U(1)$ symmetry, the theory has a $U(1)$ R-symmetry. Since the corresponding direction in the gravitational background was not involved in the solution-generating procedure, the deformed theory has ${\cal N}=1$ supersymmetry. The deformation of the supergravity background was matched to an exactly marginal operator in the field theory, thereby providing a holographic test of the methods of Leigh and Strassler \cite{Leigh}. 

For the case of ${\cal N}=4$ super Yang-Mills theory, one can perform a chain of T-duality-shift-T-duality transformations which involve the direction that corresponds to the $U(1)$ R-symmetry, which results in additional marginal deformations that do not preserve supersymmetry \cite{Frolov}. In contrast, some of the Lifshitz and Schr\"{o}dinger-like fixed points that we consider have a global symmetry group that contains $U(1)^3$, which enables us to explore various marginal deformations that preserve a minimal amount of supersymmetry.

This paper is organized as follows. In section 2, we construct the type IIB supergravity dual background that describes the marginal deformations of the effective theory of the supersymmetric lowest Landau level, which is a $(0,2)$ Landau-Ginsburg theory. In section 3, we construct the supergravity duals of deformations of field theories which preserve Schr\"odinger symmetry. The Sasaki-Einstein spaces $L^{p,q,r}$ provide countably-infinite examples of such theories that can be deformed in this manner. In section 4, we consider the marginal deformations of theories that exhibit Lifshitz scaling. The Sasaki-Einstein spaces $Y^{p,q}$ are used to construct a countably-infinite familiy of such marginally deformed theories that all have dynamical exponent $z=2$. We also consider a massive type IIA background that describes a marginally deformed theory with general dynamical exponent $z\ge 1$. Conclusions are presented in section 5.

\section{Marginal deformations of $(0,2)$ Landau-Ginsburg theory}

The bosonic sector of $U(1)^3$ truncation of $D=5$ $SO(6)$ gauged supergravity \cite{Cvetic} that keeps two neutral scalar fields $\phi_a$ has the Lagrangian
\be
e^{-1} {\mathcal L}_5=R-\fft12 (\partial\varphi_1)^2-\fft12 (\partial\varphi_2)^2+4\sum_{i=1}^3 X_i^{-1}-\fft14 \sum_{i=1}^3 X_i^{-2} (F_\2^i)^2+\fft14 \epsilon^{\mu\nu\rho\sigma\lambda} F_{\mu\nu}^1 F_{\rho\sigma}^2 A_{\lambda}^3\,,
\ee
where the two scalars are expressed in terms of three constrained scalars $X_i$ via
\be
X_1=e^{-\fft{1}{\sqrt{6}}\varphi_1-\fft{1}{\sqrt{2}} \varphi_2},\qquad X_2=e^{-\fft{1}{\sqrt{6}}\varphi_1+\fft{1}{\sqrt{2}} \varphi_2},\qquad X_3=e^{\fft{2}{\sqrt{6}}\varphi_1},
\ee
with $X_1X_2X_3=1$. Also, we are taking $g=1$. A family of magnetic AdS$_3\times\R^2$ solutions found in \cite{Almuhairi1} and studied further in \cite{Almuhairi2,Donos2} are given by\footnote{There are also families of supersymmetric magnetic AdS$_3\times H^2$ \cite{Maldacena:2000mw} and AdS$_3\times S^2$ \cite{Cucu:2003bm} solutions whose marginal deformations were obtained in \cite{Ahn:2005vk}.}
\bea\label{5d-solutions}
ds_5^2 &=& L^2\ ds_{{\rm AdS}_3}^2+dy_1^2+dy_2^2\,,\nn\\
F_\2^i &=& 2q_i\ dy_1\wedge dy_2\,,\nn\\
\varphi_1 &=& f_1\,,\qquad \varphi_2=f_2\,,
\eea
where $f_1$ and $f_2$ are constants and
\be
L^{-2} = \sum_{i=1}^3 X_i^{-1},\qquad q_i^2=X_i\,.
\ee
For $f_2=0$, these solutions reduce to a ``magnetovac" solution of Romans' $D=5$ gauged supergravity \cite{Romans1}, which can be uplifted to ten \cite{Lu:1999bw} and eleven dimensions \cite{Gauntlett1}. The case in which both $f_1$ and $f_2$ vanish is a non-supersymmetric solution of minimal gauged supergravity, which arises in the near-horizon limit of magnetic black brane solutions at zero temperature \cite{DHoker}. There is a subset of solutions which preserve supersymmetry provided that they satisfy the constraint 
\be
\sum_{i=1}^3 q_i=0\,,
\ee
as well as
\be
L^{-1}=\fft12 \sum_{i=1}^3 X_i\,,\qquad 2\sum_{i=1}^3 X_i^2=\left( \sum_{i=1}^3 X_i\right)^2.
\ee
Since the solutions degenerate if any of the $q_i$ vanish, we require all of them to be nonzero. 

It has been proposed that the above supersymmetric subset of solutions provide a supergravity dual description of the effective field theory of the supersymmetric lowest Landau level \cite{Almuhairi2}. This is a $(0,2)$ Landau-Ginsburg theory that can arise in the low-energy limit of a flow from four-dimensional ${\cal N}=4$ super Yang-Mills theory. The flow is described by a supergravity solution that smoothly interpolates from AdS$_5$ to AdS$_3\times \R^2$. The superpotential for the UV theory is given by
\be\label{UV-superpotential}
{\rm Tr} \left( \Phi_1\Phi_2\Phi_3-\Phi_1\Phi_3\Phi_2\right)\,,
\ee
where $\Phi_i$ are three chiral superfields. In the low-energy theory, there are $\nu_1=|q_1| N_{\Phi} N^2$ chiral multiplets $\Theta_{1i}$, $\nu_2=|q_2| N_{\Phi} N^2$ chiral multiplets $\Theta_{2j}$ and $\nu_3=\nu_1+\nu_2$ fermionic multiplets $\Psi_{3k}$, where $N_{\Phi}=BV_2/2\pi$, $B$ is the magnitude of the magnetic field, $V_2$ is the volume of the magnetic plane and $q_1$ and $q_2$ are the charges of the chiral multiplets. Inserting these modes into the superpotential (\ref{UV-superpotential}) yields a low-energy $(0,2)$ superpotential of the schematic form 
\be\label{superpotential1}
\left( a_{ijk}-b_{ijk}\right) \Theta_{1i} \Theta_{2j} \Psi_{3k}\,,
\ee
where $a_{ijk}$ and $b_{ijk}$ are induced by the two terms in the UV superpotential (\ref{UV-superpotential}) and are associated with the overlap of the lowest Landau level wavefunctions. One can consolidate these factors in the expression for the superpotential (\ref{superpotential1}) by taking $c_{ijk}\equiv a_{ijk}-b_{ijk}$, as was done in \cite{Almuhairi2}.

We will consider deformations of the supergravity backgrounds which leave the AdS$_3\times \R^2$ subspace intact. Such deformations can be considered to be marginal in the sense that they maintain the two-dimensional conformal symmetry of the undeformed field theory. First we lift the solutions (\ref{5d-solutions}) to ten-dimensional type IIB supergravity on a 5-sphere using the Kaluza-Klein reduction ansatz in \cite{Cvetic} to give
\bea\label{AdS-solution1}
ds_{10}^2 &=& \sqrt{\Delta}\ ds_5^2+\fft{1}{\sqrt{\Delta}}\ \sum_{i=1}^3 X_i^{-1} \left( d\mu_i^2+\mu_i^2 D\phi_i^2\right),\nn\\
F_\5 &=& L^3\ \epsilon_\3\wedge \sum_{i=1}^3 \left[ 2(X_i^2\mu_i^2-\Delta X_i)\ dy_1\wedge dy_2+q^i X_i^{-1} d(\mu_i^2)\wedge D\phi_i\right]+\mbox{dual}\,,
\eea
where $\epsilon_\3$ is the volume-form for AdS$_3$,
\be
D\phi_i= d\phi_i+A^i,\qquad \Delta=\sum_{i=1}^3 X_i \mu_i^2\,,\qquad dA^i=F_\2^i\,,
\ee
and the $\mu_i$ obey the constraint $\sum_{i=1}^3 \mu_i^2=1$. Next, in order to study deformations which preserve a $U(1)\times U(1)$ global symmetry as well as the $(0,2)$ supersymmetry, it is convenient to define the coordinates
\be\label{new-phi}
\phi_1=\a_1\psi-\td\phi_2\,,\qquad \phi_2=\a_2\psi+\td\phi_1+\td\phi_2\,,\qquad \phi_3=\a_3\psi-\td\phi_1\,,
\ee
where
\be
\a_i=\fft{q_i^2}{q_1^2+q_2^2+q_3^2}\,.
\ee
Then the Killing vector $\partial_{\psi}$ matches with the $U(1)$ superconformal R-symmetry computed through $c$-extremization \cite{Benini:2012cz,Benini:2013cda}\footnote{See \cite{Karndumri:2013dca,Jeong:2014iva} for additional examples of the supergravity dual of $c$-extremization applied to various AdS$_3$ backgrounds, including null-warped AdS$_3$ solutions obtained via TsT transformations.}. We T-dualize to type IIA theory along the $\td\phi_1$ direction and lift the resulting solution to eleven dimensions. Next, we perform the coordinate transformation 
\be
\td\phi_2\rightarrow\td\phi_2+\gamma \td\phi_1+\sigma x_{11}\,,
\ee
where $x_{11}$ is the eleventh direction. Reducing along the transformed $x_{11}$ direction and T-dualizing back along the transformed $\td\phi_1$ direction yields the deformed type IIB solution
\bea\label{section1-solution}
ds_{10}^2 &=& \fft{1}{G^{\fft14}\sqrt{\Delta}}\left[ \Delta\ ds_5^2+ \sum_{i=1}^3 X_i^{-1} \left( d\mu_i^2
+G \mu_i^2 D\phi_i^2\right)
+ (\gamma^2+\sigma^2) \fft{G}{\Delta} \prod_{i=1}^3 \fft{\mu_i^2}{X_i} \Bigg( \sum_{i=1}^3 D\phi_i\Bigg)^2\right],\nn\\
F_\5 &=& L^3\ \epsilon_\3\wedge \sum_{i=1}^3 \left[ 2(X_i^2\mu_i^2-\Delta X_i)\ dy_1\wedge dy_2+q^i X_i^{-1} d(\mu_i^2)\wedge D\phi_i\right]+\mbox{dual}\,,\nn\\
F_\3^{RR} &=& \sigma\ dB_\2 - \gamma\ GH C_\3\,,\qquad F_\3^{NS} = \gamma\ d[GH B_\2]+\sigma\ C_\3\,,\nn\\
e^{2\phi} &=& GH^2\,,\qquad \chi=-\gamma\sigma g H^{-1}\,,
\eea
where
\bea
B_\2 &=& \fft{1}{\Delta H} \sum_{j<k} (-1)^{j+k} \fft{\mu_j^2 \mu_k^2}{X_jX_k}\ D\phi_j\wedge D\phi_k\,,\\
\ast C_\3 &=& L^3\ \epsilon_\3\wedge \sum_{i=1}^3\left[ 2(X_i^2\mu_i^2-\Delta X_i)\ dy_1\wedge dy_2+
q^i X_i^{-1} d(\mu_i^2)\wedge D\phi_i\right]\wedge B_\2\,,\nn
\eea
and
\bea\label{AdS-functions1}
H &=& 1+\sigma^2 g\,,\qquad G^{-1}=1+(\gamma^2+\sigma^2) g\,,\qquad g=\fft{1}{\Delta} \sum_{j<k} \fft{\mu_j^2 \mu_k^2}{X_j X_k}\,.
\eea

We will now consider the effect of these marginal deformations on the superpotential of the dual field theory. For the marginally deformed UV theory, the superpotential is given by \cite{Leigh}
\be
{\rm Tr} \left( e^{i\pi\beta}\Phi_1\Phi_2\Phi_3 - e^{-i\pi\beta} \Phi_1\Phi_3\Phi_2\right)\,.
\ee
where $\beta=\gamma-i \sigma$. The dual supergravity description is provided by the Lunin-Maldacena background \cite{Lunin}. For the low-energy theory, this induces a marginally deformed superpotential with the schematic form
\be
\left( e^{i\pi\beta} a_{ijk}-e^{-i\pi\beta} b_{ijk}\right) \Theta_{1i} \Theta_{2j} \Psi_{3k}\,.
\ee

\section{Marginal deformations of theories with Schr\"odinger symmetry}

\subsection{An example with a five-sphere}

We will now consider gravity duals of theories which exhibit Schr\"odinger symmetry, which can be used to study non-relativistic systems such as atomic gases at ultra-low temperature \cite{Son,Balasubramanian2}. An example of such a solution in type IIB theory is given by
\cite{Herzog,Martelli}
\bea\label{null1}
ds_{10}^2 &=& r^2 \left( -2dt dy-r^2 dt^2+d\vec x^2\right)+\fft{dr^2}{r^2}+ds_{S^5}^2\,,\nn\\
F_\5 &=& 4 \left(\Omega_\5+\ast \Omega_\5\right),\nn\\
B_\2^{NS} &=& -r^2 dt\wedge \left( d\psi+A_\1\right),
\eea
where $F_\3^{NS}=dB_\2^{NS}$ and $\Omega_\5$ is the volume-form on a unit $S^5$. 
The metric on $S^5$ has been expressed as a $U(1)$ bundle over $\CP^2$:
\be
ds_{S^5}^2=\left( d\psi+A_\1\right)^2+ds_{\CP^2}^2\,,
\ee
with the metric for the $\CP^2$ base space
\be
ds_{\CP^2}=d\sigma^2+\ft14 s_{\sigma}^2 (d\theta^2+s_{\theta}^2 d\phi^2)+\ft14 s_{\sigma}^2 c_{\sigma}^2 (d\beta+c_{\theta} d\phi)^2\,,
\ee
and the K\"ahler potential
\be
A_\1=\ft12 s_{\sigma}^2 (d\beta+c_{\theta} d\phi)\,.
\ee

The solution (\ref{null1}) can be obtained by applying a null Melvin twist to the AdS$_5\times S^5$ background, which enables the dual field theory interpretation of the result to be ${\cal N}=4$ super Yang-Mills twisted by an R-charge \cite{Herzog,Martelli}. Rather than obeying conformal scale invariance, the temporal and spatial coordinates in the theory scale anisotropically. This corresponds to the metric in (\ref{null1}) obeying the scaling relation
\be
t\rightarrow \lambda^z t\,,\qquad \vec x\rightarrow \lambda \vec x\,,\qquad r\rightarrow \lambda^{-1} r\,,\qquad y\rightarrow \lambda ^{2-z} y\,,
\ee
where in this case the dynamical exponent $z=2$. If the momentum along the $y$ direction is interpreted as rest mass, then this geometry describes a system which exhibits time and space translation invariance, spatial rotational symmetry and invariance under the combined operations of time reversal and charge conjugation.

One can apply U-duality to generate deformations of the AdS$_5\times S^5$ solution using the $\beta$ and $\phi$ directions, which correspond to marginal deformations of ${\cal N}=4$ super Yang-Mills theory that preserve conformal symmetry and ${\cal N}=1$ supersymmetry \cite{Lunin}. A null Melvin twist of this deformed solution has been presented in \cite{Bobev}. Since this null Melvin twist does not involve the $\beta$ and $\phi$ directions, one can obtain the same result by generating the marginal deformations directly on the solution (\ref{null1}). The final solution can be interpreted as describing marginally deformed super Yang-Mills theory twisted by an R-charge. Although the conformal symmetry has been broken by the null Melvin twist, the deformations are marginal with respect to the Schr\"odinger symmetry.

Alternatively, one can generate marginal deformations using the $U(1)$ symmetry associated with the $y$ direction. Working directly from the solution given by (\ref{null1}), we perform T-duality along the $\beta$ direction and lift to eleven dimensions. Then we perform the coordinate transformation $y\rightarrow y+\gamma_1\beta+\gamma_2 x_{11}$, where $x_{11}$ is the eleventh direction. Upon reducing to type IIA theory along the transformed $x_{11}$ direction and T-dualizing along the transformed $\beta$ direction, we obtain the deformed type IIB solution
\bea
ds_{10}^2 &=& r^2 \left( -2dt dy-r^2 H\ dt^2+d\vec x^2\right)+\fft{dr^2}{r^2}+ds_{S^5}^2\,,\nn\\
F_\5 &=& 4 \left(\Omega_\5+\ast \Omega_\5\right),\nn\\
C_\2^{RR} &=& \gamma_2\ r^2\ dt\wedge (s_{\sigma}^2\ d\psi+A_\1)\,,\nn\\
B_\2^{NS} &=& -r^2\ dt\wedge \left[ (1+\gamma_1^2 s_{\sigma}^2) d\psi+(1+\gamma_1) A_\1\right],\nn\\
\phi &=& \chi=0\,,
\eea
where
\be
H=1+(\gamma_1+\gamma_1^2+\gamma_2^2) s_{\sigma}^2\,,
\ee
and $F_\3^{RR}=dC_\2^{RR}$. Since the Schr\"odinger portion of the geometry remains intact for constant $\sigma$ slicings, the dual field theory retains the Schr\"odinger symmetry.

In contrast to the previously-mentioned case, the procedure for obtaining these marginal deformations does not commute with the null Melvin twist, since the latter operation entails taking a lightlike boost in the $y$ direction. In fact, the order of operations has a qualitative effect on the field theory interpretation of the deformations. Namely, if the above deformations had been generated on the AdS$_5\times S^5$ background, the result would have been a nonlocal field theory \cite{Bergman}.

\subsection{Countably-infinite examples with the $L^{p,q,r}$ spaces}

The above construction can be generalized to gravity duals that involve the countably-infinite five-dimensional cohomogeneity two Sasaki-Einstein spaces $L^{p,q,r}$ \cite{page}. The metric for the $L^{p,q,r}$ spaces can be written in the canonical form
\be
ds_{L^{p,q,r}}^2=\left( d\tau+A_\1\right)^2+ds_4^2\,,
\ee
where the metric of the four-dimensional Einstein-K\"ahler base space is
\be
ds_4^2 = \fft{\rho^2 dx^2}{4\Delta_x}+\fft{\rho^2 d\theta^2}{\Delta_{\theta}}+\fft{\Delta_x}{\rho^2} \left( \fft{s_{\theta}^2}{\alpha}\ d\phi+\fft{c_{\theta}^2}{\beta}\ d\psi\right)^2+\fft{\Delta_{\theta} s_{\theta}^2 c_{\theta}^2}{\rho^2} \left( \fft{\alpha-x}{\alpha}\ d\phi-\fft{\beta-x}{\beta}\ d\psi\right)^2,
\ee
the K\"ahler potential is
\be
A_\1=\fft{\alpha-x}{\alpha} s_{\theta}^2\ d\phi+\fft{\beta-x}{\beta} c_{\theta}^2\ d\psi\,,
\ee
and the various functions are given by
\be
\Delta_x=x(\alpha-x)(\beta-x)-\mu\,,\qquad \Delta_{\theta}=\alpha c_{\theta}^2+\beta s_{\theta}^2\,,\qquad \rho^2=\Delta_{\theta}-x\,.
\ee
Details regarding the conditions that ensure that the $L^{p,q,r}$ metric extends smoothly onto a complete and non-singular manifold are given in \cite{page}. The result is that the parameters $\alpha$ and $\beta$ as well as the roots of $\Delta_x$ can be expressed in terms of coprime integer triples $p$, $q$ and $r$ which satisfy $0<p\le q$ and $0<r<p+q$, with $p$ and $q$ each coprime to $r$ and to $s=p+q-r$. The $L^{p,q,r}$ spaces have $U(1)^3$ isometry in general, which is enlarged to $SU(2)\times U(1)^2$ for $p+q=2r$, which corresponds to the subset of $Y^{p,q}=L^{p-q,p+q,p}$ spaces found in \cite{Ypq1,Ypq2}.
Note that the $L^{p,q,r}$ metric reduces to that of $S^5$ for $\mu=0$, and $\mu$ can otherwise be rescaled to $\mu=1$. 

Consider the type IIB solution
\bea\label{Labc-solution}
ds_{10}^2 &=& r^2 \left( -2dt dy-r^2 dt^2+d\vec x^2\right)+\fft{dr^2}{r^2}+ds_{L^{p,q,r}}^2\,,\nn\\
F_\5 &=& 4\left(\ep_\5+\ast\ep_\5\right),\nn\\
B_\2^{NS} &=& -r^2\ dt\wedge \left( d\tau+A_\1\right),
\eea
which can be obtained by applying a null Melvin twist to the AdS$_5\times L^{p,q,r}$ background.
$\ep_\5$ denotes the volume-form on the $L^{p,q,r}$ space. We will first consider marginal deformations which involve the $U(1)$ symmetries associated with the $\phi$ and $\psi$ directions. We T-dualize the solution (\ref{Labc-solution}) along the $\phi$ direction, lift to eleven dimensions and perform the coordinate transformation $\psi\rightarrow \psi+\gamma\phi+\sigma x_{11}$, 
where $x_{11}$ is the eleventh direction. Upon reducing and T-dualizing back to type IIB theory along the transformed $x_{11}$ and $\phi$ directions, respectively, we obtain the deformed solution
\bea
ds_{10}^2 &=& G^{-1/4} \Bigg[ r^2 \left( -2dt dy-r^2 dt^2+d\vec x^2\right)+\fft{dr^2}{r^2}+\fft{\rho^2}{4\Delta_x}\ dx^2+\fft{\rho^2}{\Delta_{\theta}}\ d\theta^2+f\ d\tau^2+ \fft{G c_{\theta}^2}{\beta^2 \rho^2 b}\ D\psi^2\nn\\
&+& \fft{G s_{\theta}^2}{\alpha^2 \rho^2 a} \left( d\phi+B_\1\right)^2\Bigg],\nn\\
F_\5 &=& \fft{2\rho^2 s_{\theta}c_{\theta}}{\alpha\beta}\ d\tau\wedge dx\wedge d\theta\wedge 
\left( G\ d\psi+\fft{\gamma}{\alpha} (\alpha-x)r^2s_{\theta}^2\ dt\right)\wedge (d\phi+B_\1)+\mbox{dual}\,,\nn\\
F_\3^{RR} &=& \fft{2\gamma}{\alpha\beta H} \rho^2 s_{\theta} c_{\theta}\ d\tau\wedge dx\wedge d\theta+
\fft{\sigma}{H} \left( \fft{\gamma g}{\alpha}\ d[(\alpha-x)r^2 s_{\theta}^2 dt]-G^{-1} d[g G\ D\psi]
\right)\wedge (d\phi+B_\1)\,,\nn\\
F_\3^{NS} &=& -d[r^2\ dt\wedge \left( d\tau+A_\1\right)]+\gamma\ d[g G\ D\psi\wedge (d\phi+B_\1)]+\fft{2\sigma}{\alpha\beta} \rho^2 s_{\theta} c_{\theta}\ d\tau\wedge dx\wedge d\theta\,,\nn\\
e^{2\phi} &=& GH^2\,,\qquad \chi=-\gamma\sigma gH^{-1}\,,
\eea
where
\bea
B_\1 &=& \alpha \rho^2 (\alpha-x) a\ d\tau-\fft{\alpha}{\beta} a c_{\theta}^2\ d\psi-\fft{\gamma}{\beta} e r^2 c_{\theta}^2\ dt\,,\nn\\
D\psi &=& d\psi+\beta\rho^2 be\ d\tau+\gamma r^2 s_{\theta}^2\left(\fft{\alpha -x}{\alpha}\right) dt\,,
\eea
\be
H = 1+\sigma^2 g\,,\qquad G^{-1} = 1+(\gamma^2+\sigma^2)g\,,\qquad g = \fft{s_{\theta}^2 c_{\theta}^2}{\alpha^2\beta^2\rho^4 ab}\,,
\ee
and we have defined the functions
\bea
a^{-1} &=& \alpha (\alpha -x)\rho^2-\mu s_{\theta}^2\,,\nn\\
b^{-1} &=& \beta (\beta -x)\rho^2-\mu c_{\theta}^2-a s_{\theta}^2 c_{\theta}^2\,,\nn\\
e &=& \beta-x+(\alpha-x)a s_{\theta}^2\,,\nn\\
f &=& 1-(\alpha-x)^2 \rho^2 a s_{\theta}^2-\rho^2 b e^2 c_{\theta}^2\,.
\eea

Since the above procedure for generating marginal deformations commutes with the null Melvin twist, the same result can be obtained by applying the null Melvin twist to the marginal deformations of AdS$_5\times L^{p,q,r}$ that are described in \cite{Ahn}. This enables us to interpret the final solution as describing marginal deformations of the corresponding quiver gauge theories twisted by an R-charge, for which conformal symmetry is broken but the Schr\"odinger symmetry is preserved. 

The field theories have $p+3q$ chiral fields which come in six different types: $q$ $Y$, $(p+q-r)$ $U_1$, $r$ $U_2$, $p$ $Z$, $(r-p)$ $V_1$ and $(q-r)$ $V_2$ fields. For the $Y^{p,q}$ subset, the $U_i$ fields become a doublet under the $SU(2)$ flavor symmetry. For the undeformed $L^{p,q,r}$ theories, a superpotential can be built out of these fields which has the following schematic form \cite{Franco}:
\be
W=2p\ \Tr(YU_1ZU_2)+2(q-r)\ \Tr(YU_1V_1)+2(r-p)\ \Tr(YU_2V_2)\,.
\ee
For the above marginal deformations, the quartic portion of the superpotential is altered as follows:
\be
\Tr(YU_1ZU_2-YU_2ZU_1)\rightarrow \Tr\left( e^{i\pi\td\beta}YU_1ZU_2-e^{-i\pi\td\beta}YU_2ZU_1\right),
\ee
where the complex deformation parameter\footnote{We put a tilde in order to distinguish this from the parameter $\beta$ in the $L^{p,q,r}$ metric.} $\td\beta=\gamma-i \sigma$. This interpretation of the marginal deformations survives taking the null Melvin twist.

Now we consider marginal deformations which involve the $U(1)$ symmetries associated with the $y$ and $\phi$ directions. We T-dualize along the $\phi$ direction, lift to eleven dimensions and perform the coordinate transformation $y\rightarrow y+\gamma_1 \phi+\gamma_2 x_{11}$. Upon reducing to type IIA theory along the transformed $x_{11}$ direction and T-dualizing along the transformed $\phi$ direction, we obtain the deformed type IIB solution
\bea
ds_{10}^2 &=& r^2 \left( -2dt dy-r^2 H\ dt^2+d\vec x^2\right)+\fft{dr^2}{r^2}+ds_{L^{p,q,r}}^2\,,\nn\\
F_\5 &=& 4\left(\ep_\5+\ast\ep_\5\right),\nn\\
C_\2^{RR} &=& \gamma_2\ \fft{r^2 s_{\theta}^2}{\alpha^2 \rho^2 b}\ dt\wedge (d\phi+B_\1)\,,\nn\\
B_\2^{NS} &=& -r^2\ dt\wedge \left( \left( d\tau+A_\1\right)+\fft{\gamma_1 s_{\theta}^2}{\alpha^2 \rho^2 b}  (d\phi+B_\1)\right),\nn\\
\phi &=& \chi=0\,,
\eea
where
\be
B_\1 = \fft{\alpha \rho H}{\beta} \left( \beta (\alpha-x)\rho\ d\tau-c_{\theta}^2\ d\psi\right),
\ee
and
\be
H=1+s_{\theta}^2 \left[ \fft{\gamma_1^2+\gamma_2^2}{\alpha^2 \rho^2 b}+2\gamma_1 \left( \fft{x-\alpha}{\alpha}\right)\right].
\ee
For constant slicings of $\theta$ and $x$, the Schr\"odinger portion of the geometry remains intact, and so the dual field theory still has Schr\"odinger symmetry.

\section{Marginal deformations of Lifshitz vacua}

\subsection{Lifshitz-Chern-Simons gauge theories}

We will now consider some examples of gravity dual descriptions of marginal deformations of field theories with Lifshitz scaling \cite{Kachru}. A candidate gravity dual for a class of $2+1$ dimensional Lifshitz-Chern-Simons field theories with dynamical exponent $z=2$ was constructed in \cite{Balasubramanian}. The type IIB solution is given by
\bea\label{Lifshitz1}
ds_{10}^2 &=& r^2 (2dt dx_3+d\vec x^2)+f\ dx_3^2+\fft{dr^2}{r^2}+ds_{S^5}^2\,,\nn\\
F_\5 &=& 2r^3\ d^4x\wedge dr+\mbox{dual}\,,\nn\\
\chi &=& \fft{Qx_3}{L_3}\,,\qquad \phi=0\,,
\eea
where
\be
f=\fft{Q^2}{4L_3^2}-\fft{r_0^4}{r^2}\,,
\ee
and $d\vec x^2=dx_1^2+dx_2^2$. It has been proposed that this background describes non-Abelian Lifshitz Chern-Simons gauge theories which can be realized as deformations of DLCQ ${\cal N}=4$ super Yang-Mills theory. A Chern-Simons term explicitly breaks parity and time-reversal symmetries. The background (\ref{Lifshitz1}) asymptotically exhibits the following scaling symmetry:
\be
t\rightarrow\lambda^2 t\,,\qquad \vec x\rightarrow \lambda \vec x\,,\qquad r\rightarrow \lambda^{-1} r\,,\qquad x_3\rightarrow x_3\,.
\ee
$x_3$ is a compact direction and therefore cannot scale, since scaling would change the compactification radius. The $x_3$ circle shrinks to zero size at
\be
r=r_{\star}=\fft{2L_3 r_0^2}{Q}\,.
\ee
The metric is geodesically incomplete in the region $r\ge r_{\star}$, and a straightforward way of extending geodesics past $r_{\star}$ leads to closed timelike curves \cite{Copsey}. The implications of this hidden singularity on the dual field theory as well as its resolutions are discussed in \cite{Balasubramanian}.

The metric on $S^5$ can be expressed as
\be
ds_{S^5}^2=\sum_{i=1}^3 \left( d\mu_i^2+\mu_i^2 d\phi_i^2\right),
\ee
where $\sum_{i=1}^3 \mu_i^2=1$. Before applying the solution-generating technique, we perform the coordinate transformation (\ref{new-phi}). We T-dualize along the $x_3$ direction to obtain a solution in massive type IIA theory and perform the transformation
\be
\td\phi_1\rightarrow \td\phi_1+\gamma_1 x_3\,.
\ee
Then we T-dualize back to type IIB theory along the transformed $x_3$ direction to obtain the deformed type IIB solution
\bea\label{Lifshitz1-deformed}
ds_{10}^2 &=& \fft{1}{G^{1/4}} \Bigg[ -\fft{r^4}{f}\ dt^2+r^2 d\vec x^2+\fft{dr^2}{r^2}+G f \left( dx_3+\fft{r^2}{f}\ dt\right)^2+ds_{S^5}^2
- \gamma_1^2 f G\  d\phi_{23}^2\Bigg],\nn\\
F_\5 &=& 2r^3\ d^4x\wedge dr+\mbox{dual}\,,\nn\\
F_\3^{RR} &=& \gamma_1 \fft{QGr^2}{L_3}\ dt\wedge dx_3\wedge d\phi_{23}\,,\qquad
F_\3^{NS} = -\gamma_1\ d\left[G \left( f\ dx_3+r^2 dt\right)
\wedge d\phi_{23}\right],\nn\\
e^{2\phi} &=& G\,,\qquad \chi = \fft{Qx_3}{L_3}\,,
\eea
where
\be
d\phi_{23}\equiv \mu_2^2\ d\phi_2-\mu_3^2\ d\phi_3\,,\qquad G^{-1} = 1+\gamma_1^2 f (\mu_2^2+\mu_3^2)\,.
\ee
While the solution (\ref{Lifshitz1}) has a null Killing vector, the deformed solution (\ref{Lifshitz1-deformed}) has a timelike Killing vector $\partial_t$ generating the time translations of the dual field theory for $r>0$ and $\mu_2^2+\mu_3^2>0$.

A different deformation can be obtained by T-dualizing along the $\td\phi_1$ direction to type IIA theory, performing the transformation
\be
\td\phi_2\rightarrow \td\phi_2+\gamma_2 \td\phi_1\,,
\ee
and T-dualizing back to type IIB theory along the transformed $\td\phi_1$ direction. The resulting deformed type IIB solution is given by
\bea
ds_{10}^2 &=& \fft{1}{G^{1/4}} \Bigg[ r^2 (2 dt dx_3+d\vec x^2)+f\ dx_3^2+\fft{dr^2}{r^2}+\sum_{i=1}^3 \left( d\mu_i^2+G\mu_i^2 d\phi_i^2\right)
+ \gamma_2^2 G \prod_{j=1}^3 \mu_j^2 \Big(\sum_{k=1}^3 d\phi_k\Big)^2\Bigg],\nn\\
F_\5 &=& 2r^3\ d^4x\wedge dr+\mbox{dual}\,,\nn\\
F_\3^{RR} &=& \gamma_2\ \fft{QG}{L_3}\ dx_3\wedge 
\sum_{j<k} (-1)^{j+k} \mu_j^2 \mu_k^2\ d\phi_j\wedge d\phi_k
+\fft32 \gamma_2\ d(\mu_2^2)\wedge d(\mu_3^2)\wedge d\psi\,,\nn\\
F_\3^{NS} &=& \gamma_2\ d\left[G  
\sum_{j<k} (-1)^{j+k} \mu_j^2 \mu_k^2\ d\phi_j\wedge d\phi_k
\right],\\
e^{2\phi} &=& G\,,\qquad \chi = \fft{Qx_3}{L_3}\,,\nn
\eea
where
\be
G^{-1} = 1+\gamma_2^2 \sum_{j<k} \mu_j^2 \mu_k^2\,.
\ee
Note that the Killing vector $\partial_t$ remains null for this deformation. If the $\gamma_1$ and $\gamma_2$ deformations are turned on at the same time, then the fiber structure of the resulting solution implies that the $x_3$ direction is periodic. However, we will restrict ourselves to the scenario in which the $x_3$ direction is extended.

\subsection{Countably-infinite Lifshitz vacua with dynamical exponent $z=2$}

Infinite families of Lifshitz solutions of ten and eleven-dimensional supergravity with dynamical exponent $z=2$ were constructed in \cite{Donos}. As the starting point for applying the solution-generating technique, we will consider solutions in eleven-dimensional supergravity. Before turning to a countably infinite family of solutions involving the $Y^{p,q}$ spaces, we first consider the eleven-dimensional solution involving the space $T^{1,1}$, which is given by \cite{Donos}
\bea
ds_{11}^2 &=& ds_4^2+\left( d\sigma+\fft{1}{\sqrt{18}} (c_1 d\phi_1-c_2 d\phi_2)\right)^2
+ dx_{11}^2 + \fft19 (d\psi-c_1 d\phi_1-c_2 d\phi_2)^2\nn\\
&+& \fft16 (d\theta_1^2+s_1^2 d\phi_1^2+d\theta_2^2+s_2^2 d\phi_2^2)\,,\nn\\
G_\4 &=& dt\wedge d\left[ r^4\ dx_1\wedge dx_2+r^2\ dx_{11}\wedge \left( d\sigma+\fft{1}{\sqrt{18}} (c_1 d\phi_1-c_2 d\phi_2)\right)\right],
\eea
where 
\be
ds_4^2=-r^4\ dt^2+r^2 (dx_1^2+dx_2^2)+\fft{dr^2}{r^2}\,,
\ee
and we denote $c_1\equiv\cos\theta_1$, $s_2\equiv \sin\theta_2$, etc. We perform the coordinate transformation
\be
\phi_2\rightarrow \phi_2+\gamma x_{11}\,,
\ee
reduce to type IIA theory along the transformed $x_{11}$ direction and T-dualize along the transformed $\sigma$ direction to obtain the type IIB solution
\bea\label{final-Lifshitz1}
ds_{10}^2 &=& G^{-1/4} \Bigg[ ds_4^2+\fft16 \left(d\theta_1^2+s_1^2 d\phi_1^2+d\theta_2^2+GK s_2^2 d\phi_2^2\right) + \fft{G}{9K} \left( d\psi-c_1 d\phi_1-Kc_2 d\phi_2\right)^2\nn\\
&+& G(d\sigma+r^2 dt)^2\Bigg]
,\nn\\
F_\5 &=& 4r^3\ d\sigma\wedge dt\wedge dr\wedge dx_1\wedge dx_2
 -\gamma H r^2 \left( 1+\ft{1}{18} \gamma G c_2^2\right) d\sigma\wedge dt\wedge C_\2\wedge B_\1
+\mbox{dual}\,,\nn\\
F_\3^{RR} &=& -\gamma H\ d\left[ B_\1\wedge (d\sigma+r^2 dt)\right]
-\ft{1}{\sqrt{18}}\gamma H c_2\ C_\2\wedge d\sigma
+\ft12 H G\ dG^{-1}\wedge B_\1\wedge (d\sigma +r^2 dt)\,,
\nn\\
F_\3^{NS} &=& C_\2\wedge d\sigma-\ft{1}{\sqrt{18}} \gamma\ d\left[ G c_2\ B_\1\wedge (d\sigma+r^2 dt)\right],
\nn\\
e^{2\phi} &=& GH^{-2}\,,\qquad \chi=\ft{1}{\sqrt{18}} \gamma Hc_2\,,
\eea
where
\bea
B_\1 &=& \ft19 c_2 (d\psi-c_1 d\phi_1-c_2 d\phi_2)-\ft16 s_2 d\phi_2\,,\nn\\
C_\2 &=& \ft{1}{\sqrt{18}} (s_2d\theta_2\wedge d\phi_2-s_1d\theta_1\wedge d\phi_1)\,,
\eea
and
\be
G^{-1}=1+\ft{\gamma^2}{9}+\ft{\gamma^2}{18} s_2^2\,,\qquad H^{-1} = 1+\ft{\gamma^2}{6}\,,\qquad K^{-1}=1+\ft{\gamma^2}{6} s_2^2\,.
\ee
Note that while the Killing vector $\partial_t$ is null for $\gamma=0$, it is timelike for $\gamma\neq 0$ and $r>0$.

The above construction can be generalized to gravity duals that involve the Sasaki-Einstein spaces $Y^{p,q}$, which are characterised by two coprime positive integers $p$ and $q$ with $q<p$ \cite{Ypq1,Ypq2}. We begin with the eleven-dimensional solution \cite{Donos}
\bea
ds_{11}^2 &=& f^{1/3} ds_4^2+f^{-2/3} \left[ \left( d\sigma-\fft{D\beta}{\sqrt{72} (1-y)}\right)^2+dx_{11}^2\right]\nn\\
&+& f^{1/3} \left[ \fft19 \left( d\psi+y\ D\beta-c_{\theta} d\phi\right)^2+\fft{1-y}{6} (d\theta^2+s_{\theta}^2d\phi^2)+\fft{dy^2}{g}+\fft{g}{36}\ D\beta^2\right],\nn\\
G_\4 &=& dt\wedge d\left[ r^4\ dx_1\wedge dx_2+\fft{r^2}{f}\ dx_{11}\wedge \left( d\sigma-\fft{D\beta}{\sqrt{72} (1-y)}
\right)\right],
\eea
where 
\be
ds_4^2=-\fft{r^4}{f}\ dt^2+r^2 (dx_1^2+dx_2^2)+\fft{dr^2}{r^2}\,,
\ee
\be
D\beta=d\beta+c_{\theta} d\phi\,,\qquad g=\fft{2(a-3y^2+2y^3)}{1-y}\,,
\ee
and $c_{\theta}\equiv\cos\theta$, $s_{\theta}\equiv \sin\theta$. We have expressed the metric of the $Y^{p,q}$ subspace in canonical form as a $U(1)$ bundle over an Einstein-K\"ahler metric. The function $f$ satisfies
\be
-4f+\fft{2}{1-y}\ \partial_y \left[ (a-3y^2+2y^3)\ \partial_y f\right]+\fft{1}{(1-y)^4}=0\,.
\ee

Upon performing the coordinate transformation
\be
\beta\rightarrow \beta+\gamma x_{11}\,,
\ee
reducing to type IIA theory along the transformed $x_{11}$ direction and T-dualizing along the transformed $\sigma$ direction, we obtain the type IIB solution
\bea
ds_{10}^2 &=& G^{-1/4} \Bigg[ ds_4^2+\fft{1-y}{6}\ (d\theta^2+s_{\theta}^2 d\phi^2)+\fft{dy^2}{g} +
\fft{Kg}{36}\ D\beta^2+\fft{G}{9K} \left( D\psi-\ft{\gamma^2}{36} K fgy\ D\beta\right)^2\nn\\
&+& fG \left( d\sigma+\fft{r^2}{f} dt\right)^2\Bigg]
,\nn\\
F_\5 &=& 4r^3\ d\sigma\wedge dt\wedge dr\wedge dx_1\wedge dx_2
+ \gamma\ \fft{Gr^2}{f}\ d\sigma\wedge dt\wedge dB_\1\wedge C_\1+\mbox{dual}\,,
\nn\\
F_\3^{RR} &=& \gamma\ d\left[ \fft{H}{\sqrt{72} (1-y)}\right]\wedge A_\1\wedge \left( d\sigma+\fft{r^2}{f}\ dt\right)+\gamma\ d\left[ H\ C_\1\wedge \left( d\sigma+\fft{r^2}{f}\ dt\right)\right]\nn\\
&-& \gamma\ \fft{H}{\sqrt{72} (1-y)}\ dB_\1\wedge d\sigma\,,
\nn\\
F_\3^{NS} &=& d\left[ B_\1\wedge d\sigma+A_\1\wedge \left( d\sigma+\fft{r^2}{f}\ dt\right)\right],\nn\\
e^{2\phi} &=& GH^{-2}\,,\qquad \chi=\fft{\gamma H}{\sqrt{72} (1-y)}\,,
\eea
where
\be
A_\1 = \fft{\gamma^2 G}{\sqrt{72} (1-y)}\ C_\1\,,\qquad B_\1=-\fft{D\beta}{\sqrt{72} (1-y)}\,,\qquad C_\1=\fft{fy}{9}\ D\psi+\fft{fg}{36}\ D\beta\,,
\ee
\be
D\sigma=d\sigma+B_\1\,,\qquad D\psi=d\psi+y\ D\beta-c_{\theta} d\phi\,,
\ee
and
\be
K^{-1}=1+\ft{\gamma^2}{36} fg\,,\qquad G^{-1} = K^{-1}+\ft{\gamma^2}{9} f y^2\,,\qquad
H^{-1} = G^{-1}+\fft{\gamma^2}{72 (1-y)^2}\,.
\ee
The Killing vector $\partial_t$ is null for $\gamma=0$ and timelike for $\gamma\neq 0$ and $r>0$.

Alternatively, one can perform the coordinate transformation
\be
\phi\rightarrow \phi+\gamma x_{11}\,.
\ee
Then reducing along the transformed $x_{11}$ direction and T-dualizing along the transformed $\sigma$ direction yields the type IIB solution
\bea
ds_{10}^2 &=& G^{-1/4} \Bigg[ ds_4^2+\fft{1-y}{6} (d\theta^2+L s_{\theta}^2 d\phi^2)+\fft{dy^2}{g} +\fft{Kg}{36L} \left( D\beta-\ft{\gamma^2}{6} Lf(1-y)s_{\theta}^2 c_{\theta} d\phi\right)^2\nn\\
&+& \fft{G}{9K} \left[ D\psi+\ft{\gamma^2}{36} Kf(1-y) \left( gc_{\theta} D\beta+6(1-y) s_{\theta}^2 d\phi\right)\right]^2
+ Gf \left( d\sigma+\fft{r^2}{f}\ dt\right)^2\Bigg]
,\nn\\
F_\5 &=& 4r^3\ d\sigma\wedge dt\wedge dr\wedge dx_1\wedge dx_2
- \fft{\gamma}{9} G r^2 (1-y) c_{\theta}\ d\sigma\wedge dt\wedge dB_\1\wedge {\mathcal D}\psi+\mbox{dual}\,,\nn\\
F_\3^{RR} &=& \gamma\ d\left[ \fft{H c_{\theta}}{\sqrt{72} (1-y)}\right]\wedge A_\1\wedge \left( d\sigma+\fft{r^2}{f}\ dt\right) -\fft{\gamma}{9}\ d\left[ Hf(1-y)c_{\theta} {\mathcal D}\psi\wedge \left( d\sigma+\fft{r^2}{f}\ dt\right)\right]
\nn\\
&-& \gamma\ \fft{H c_{\theta}}{\sqrt{72} (1-y)}\ dB_\1\wedge d\sigma\,,\nn\\
F_\3^{NS} &=& d\left[ B_\1\wedge d\sigma+A_\1\wedge \left( d\sigma+\fft{r^2}{f}\ dt\right)\right],\nn\\
e^{2\phi} &=& GH^{-2}\,,\qquad \chi=\fft{\gamma H c_{\theta}}{\sqrt{72} (1-y)}\,,
\eea
where now we take
\be
A_\1=-\fft{\gamma^2 Gfc_{\theta}^2}{9\sqrt{72}}\ {\mathcal D}\psi\,,\qquad B_\1=-\fft{D\beta}{\sqrt{72}(1-y)}\,,
\ee
\be
{\mathcal D}\psi=D\psi-\fft{g\ D\beta}{4(1-y)}-\fft{3s_{\theta}^2}{2c_{\theta}}\ d\phi\,,
\ee
and we are now defining
\bea
L^{-1} &=& 1+\ft{\gamma^2}{6} f(1-y) s_{\theta}^2\,,\qquad K^{-1}=L^{-1}+\ft{\gamma^2}{36} fgc_{\theta}^2\,,\nn\\ 
G^{-1} &=& K^{-1}+\ft{\gamma^2}{9} f(1-y)^2 c_{\theta}^2\,,\qquad H^{-1}=G^{-1}+\fft{\gamma^2 c_{\theta}^2}{72 (1-y)^2}\,.
\eea
As with the previous deformation, the Killing vector $\partial_t$ is null for $\gamma=0$ and timelike for $\gamma\neq 0$ and $r>0$.

\subsection{An example with general dynamical exponent}

Lifshitz solutions of Romans' six-dimensional gauged, massive, ${\cal N}=4$ supergravity \cite{Romans-six} were found in \cite{Gregory}. These solutions have  with general dynamical exponent $z\ge 1$ and break supersymmetry. The geometry is a direct product of a four-dimensional Lifshitz geometry and a two-dimensional hyperboloid. The metric is given by
\be
ds_6^2 = L^2 \left( -r^{2z} dt^2+r^2 (dx_1^2+dx_2^2)+\fft{dr^2}{r^2}\right)+a^2\ dH_2^2\,,
\ee
where the metric for a hyperboloid $dH_2^2$ can be made compact by modding out a non-compact discrete subgroup of the isometry group. There is a topological restriction on $z$ in terms of the 
gauge coupling $g$ and the mass parameter $m$ of the six-dimensional theory, due to flux quantization on the compact hyperbolic space.

These solutions can be lifted to massive type IIA theory by using the consistent reduction ansatz given in \cite{massiveIIA-KK}\footnote{These solutions have also been lifted to type IIB theory using a reduction ansatz generated via non-Abelian T-duality \cite{Jeong:2013jfc}.}. The resulting ten-dimensional solution is given by\footnote{We have included terms in the form fields associated with a 2-form $B_\2$ in the six-dimensional solution that are missing in the uplifted solution presented in \cite{Gregory}.}
\bea\label{typeIIA-solution}
ds_{10}^2 &=& S^{1/12} k_0^{1/8} \Delta^{3/8} \left[ ds_6^2+k_1\ d\rho^2+k_2\ \Delta^{-1} C^2\ [d\theta^2+s_{\theta}^2 d\phi^2+(d\psi+c_{\theta} d\phi-gA_\1)^2]\right],\nn\\
F_\4 &=& k_3\ S^{1/3} C^3 \Delta^{-2} U s_{\theta}\ d\rho\wedge d\theta\wedge d\phi\wedge (d\psi-gA_\1)+k_4\, \beta a^2 L r^z S^{1/3} C\ dt\wedge H_\2 \wedge d\rho\nn\\
&+& k_5\ S^{1/3} C\ G_\2\wedge (d\psi+c_{\theta} d\phi-g A_\1)\wedge d\rho
+k_6\ S^{4/3} C^2 \Delta^{-1} s_{\theta}\ G_\2\wedge d\theta\wedge d\phi\nn\\
&-& \frac{k_7 M}{2^{3/4} k_0^2}S^{4/3}\ast _6 B_{(2)} \,,\nn\\
F_\3 &=& k_7\ \beta L^3 r S^{2/3}\ dx_1\wedge dx_2\wedge dr +\frac 23 k_7\, S^{-1/3}C B_{(2)}\wedge d\rho\,,\nn\\ 
F_\2 &=& 2^{-3/4}k_7\, S^{2/3} M B_{(2)} \,,\nn\\
e^{2\phi} &=& k_0^{-5/2} S^{-5/3} \Delta^{1/2}\,,
\eea
where
\bea
 dA_\1 &=& G_\2 =\alpha L^2 r^{z-1}\ dt\wedge dr+\gamma a^2\ H_\2\,,\nn\\
B_{(2)} &=& \frac {\beta}{2}L^3 r^2 dx_1\wedge dx_2\,,
\eea
$H_\2$ is the volume-form of a unit hyperboloid, and the type IIA mass parameter is given by
\be
M=\left( \fft{2mg^3}{27}\right)^{1/4}.
\ee
The various functions are given by
\bea
\Delta &=& k_0\ C^2+k_0^{-3} S^2\,,\qquad U = k_0^{-6} S^2-3k_0^2\ C^2+4k_0^{-2} C^2-6k_0^{-2}\,,\nn\\
C &=& \cos\rho\,,\qquad S=\sin\rho\,,\qquad c_{\theta}=\cos\theta\,,\qquad s_{\theta}=\sin\theta\,,
\eea
and the various constants are
\bea
k_0 &=& e^{\phi_0/\sqrt{2}} \left( \fft{g}{3m}\right)^{1/4},\quad k_1=\fft{8}{3mg}\ e^{\sqrt{2} \phi_0},\quad k_2=\fft{2}{g^2} \left( \fft{g}{3m}\right)^{1/4} e^{-\phi_0/\sqrt{2}},\quad k_3 = -\fft{4\sqrt{2}}{3g^3} \left( \fft{g}{3m}\right)^{3/4},\nn\\ k_4 &=& 3g^2\ e^{2\sqrt{2}\phi_0}\ k_3\,,\qquad k_5=3g\ k_3\,,\qquad k_6=-\fft{\sqrt{8}}{g^2}\ e^{-3\phi_0/\sqrt{2}}\,,\qquad k_7=\sqrt{\fft{12m}{g}}\,,
\eea
and
\bea
L^2 \beta^2\ e^{\sqrt{8} \phi_0} &=& z-1\,,\nn\\
\alpha^2 &=& \gamma^2 (z-1)\,,\nn\\
L^2\gamma^2\ e^{-\sqrt{2} \phi_0} &=& \fft{(2+z)(z-3)\pm 2 \sqrt{2(z+4)}}{2z}\,,\nn\\
L^2 g^2\ e^{\sqrt{2} \phi_0} &=& 2z(4+z)\,,\nn\\
\ft12 L^2 m^2\ e^{-3\sqrt{2} \phi_0} &=& \fft{6+z\mp 2\sqrt{2(z+4)}}{z}\,,\nn\\
\fft{L^2}{a^2} &=& 6+3z\mp 2\sqrt{2(z+4)}\,.
\eea
Since the metric in (\ref{typeIIA-solution}) is singular at $\rho=0$ and $\pi$ and the string coupling diverges there, we will consider this solution away from these regions.

T-dualizing to type IIB theory along the $\psi$ direction using the extended T-duality transformation rules in \cite{massive-Tduality}, performing the transformation
\be
\phi\rightarrow \phi+\sigma \psi\,,
\ee
and T-dualizing back to massive type IIA theory along the transformed $\psi$ direction yields the deformed solution
\bea
ds_{10}^2 &=& G^{-1/4}S^{1/12} k_0^{1/8} \Delta^{3/8} \left[ ds_6^2+k_1\ d\rho^2+k_2\ \Delta^{-1} C^2 [d\theta^2+s_{\theta}^2 d\phi^2+G(d\psi+c_{\theta} d\phi-gA_\1)^2]\right],\nn\\
F_\4 &=& k_3\ G\ S^{1/3} C^3 \Delta^{-2} U s_{\theta}\ d\rho\wedge d\theta\wedge d\phi\wedge (d\psi-gA_\1)+k_4\, \beta a^2 L r^z S^{1/3} C\ dt\wedge H_\2 \wedge d\rho
\nn\\
&+& k_5\ S^{1/3} C\ G_\2\wedge (d\psi+c_{\theta} d\phi-g A_\1)\wedge d\rho
+k_6\ S^{4/3} C^2 \Delta^{-1} s_{\theta}\ G_\2\wedge d\theta\wedge d\phi
\nn\\
&-& k_7\, \frac{M}{2^{3/4} k_0^2}\, S^{4/3}\ast _6 B_{(2)}
+ \fft{\sigma k_2^{5/4} k_4\, \beta L^3 r s_{\theta}}{k_0 \Delta^{3/4} S^{1/3}}\ \sqrt{\fft{GC^7}{k_1}}
\ dx_1\wedge dx_2\wedge d\theta\wedge dr\nn\\
&-& \fft{\sigma G k_7 k_2^2 M C^4 s_{\theta}^2}{2^{3/4} k_0 \Delta}\ B_\2\wedge d\phi \wedge (d\psi-gA_{(1)})\,,\nn\\
F_\3 &=&  k_7\ \beta L^3 r S^{2/3}\ dx_1\wedge dx_2\wedge dr 
+\ft23 k_7\ S^{-1/3}C B_{(2)}\wedge d\rho
+ d\Big[ \fft{\sigma Gk_2^2 C^4 s_{\theta}^2}{\Delta S^{2/3} k_0}\ d\phi\wedge (d\psi-gA_\1)\Big],\nn\\
F_\2 &=& 2^{-3/4}k_7\, S^{2/3} M B_{(2)} +\fft{\sigma G k_2^2 M C^4 s_{\theta}^2}{k_0\Delta S^{2/3}}\ d\phi\wedge (d\psi-gA_\1)
- \sigma k_3 S^{1/3} C^3 \Delta^{-2} U s_{\theta}\ d\theta\wedge d\rho\,,\nn\\
e^{2\phi} &=& G\ k_0^{-5/2} S^{-5/3} \Delta^{1/2} \,,
\eea
where
\be
G^{-1}=1+\sigma^2\ \fft{k_2^2 C^4 s_{\theta}^2}{k_0S^{2/3} \Delta}\,.
\ee

\section{Conclusions}

A solution-generating technique based on U-duality has been used to construct supergravity backgrounds that holographically describe the marginal deformations of various non-relativistic field theories which preserve a $U(1)\times U(1)$ global symmetry. For a $(0,2)$ Landau-Ginsburg theory describing the supersymmetric lowest Landau level, we have proposed that the marginal deformations are associated with the introduction of a phase in the $(0,2)$ superpotential. This can arise in the low-energy limit of a flow from marginal deformations of four-dimensional ${\cal N}=4$ super Yang-Mills theory described by the Lunin-Maldacena background \cite{Lunin}. 

We have generated the supergravity duals of marginally deformed field theories with Schr\"odinger symmetry or Lifshitz scaling. This includes a class of Lifshitz-Chern-Simons gauge theories, as well as  
countably-infinite families of field theories whose dual gravity description involves the Sasaki-Einstein spaces $Y^{p,q}$ and $L^{p,q,r}$. These theories all have dynamical exponent $z=2$. We have also considered massive type IIA backgrounds which are dual to marginal deformations of Lifshitz theories with general dynamical exponent $z\ge 1$.

With the exception of the last example, we have focused on marginal deformations that preserve supersymmetry, these constructions can be straightforwardly generalized to scenarios in which supersymmetry is not preserved. For instance, one could begin with a supersymmetric gravity background and perform a chain of T-duality-shift-T-duality transformations to generate multiple deformations which do not preserve supersymmetry. If this is done for the case of the $(0,2)$ Landau-Ginsburg theory, then one obtains the description of a non-supersymmetric system that arises in the low-energy limit of a flow from the $6+2$-parameter deformation of super Yang-Mills theory found in \cite{Frolov}. Another possibility is to consider cases for which the undeformed field theory itself does not preserve any supersymmetry. For instance, one could study gravity duals of field theories with Schr\"odinger symmetry or Lifshitz scaling that involve five-dimensional Einstein spaces which are not Sasakian. Examples of such spaces include the $T^{p,q}$ spaces, as well as the $\Lambda^{p,q,r}$ spaces which encompass the Sasaki-Einstein spaces $L^{p,q,r}$ \cite{page2}. 

While examples without supersymmetry are more realistic, one then has less control over their behavior and must worry about instabilities. However, there are cases of non-supersymmetric string states that may be completely stable. For example, the gravity dual description for the effective field theory of the lowest Landau level presented in \cite{Almuhairi2} has a regime of parameter space in which all known instabilities are apparently absent.

%%%%%%%%%%%%%%%%%%%%%%%%%%%%%%%%%%%%%%%%
\section*{Acknowledgments}

We are grateful to Philip Argyres, Nabil Iqbal, James Liu, Oleg Lunin and Eoin \'O Colg\'ain for useful discussions. J.F.V.P. acknowledges the hospitality of the KITP, where this work was initiated. This work was supported in part by NSF grant PHY-0969482.
%%%%%%%%%%%%%%%%%%%%%%%%%%%%%%%%%%%%%%%%

\end{document}